\documentclass{aa}
\usepackage{txfonts}
\usepackage{graphicx}
\usepackage{lscape}

\usepackage{color}

\begin{document}

\title{Topological changes in the magnetic field of LQ Hya during an activity minimum
\thanks{Based on observations made with the HARPSpol instrument on
the ESO 3.6 m telescope at La Silla (Chile), under the programme IDs 084.D-0338, 086.D-0240 and 0100.D-0176 .}
\fnmsep
\thanks{The Johnson B- and V-band differential photometry and the numerical time series analysis results are available in electronic form at the CDS via anonymous ftp to \texttt{cdsarc.u-strasbg.fr} (130.79.128.5) or via \texttt{http://cdsweb.u-strasbg.fr/cgi-bin/qcat?J/A+A/}}}
\author{J.J.~Lehtinen$^{1,2}$, M.J.~K{\"a}pyl{\"a}$^{3,4,5}$, T.~Hackman$^2$, O.~Kochukhov$^6$, T.~Willamo$^2$, S.C.~Marsden$^7$, S.V.~Jeffers$^4$, G.W.~Henry$^8$, L.~Jetsu$^2$}
\titlerunning{LQ Hya magnetic field topology}
\authorrunning{J.J.~Lehtinen, M.J.~K{\"a}pyl{\"a}, T.~Hackman et al.}

\institute{
Finnish Centre for Astronomy with ESO (FINCA), University of Turku, Vesilinnantie 5, FI-20014 University of Turku, Finland
\and Department of Physics, P.O.~Box 64, FI-00014, University of Helsinki, Finland
\and Department of Computer Science, Aalto University, PO Box 15400, FI-00076 Aalto, Finland
\and Max Planck Institute for Solar System Research, Justus-von-Liebig-Weg, 3, D-37077 G\"ottingen, Germany
\and Nordita, KTH Royal Institute of Technology and Stockholm University, Hannes Alfv\'ens v\"ag 12, SE-10691 Stockholm, Sweden
\and Department Physics and Astronomy, Uppsala University, Box 516, 751 20 Uppsala, Sweden
\and University of Southern Queensland, Centre for Astrophysics, Toowoomba, QLD 4350, Australia
\and Center of Excellence in Information Systems, Tennessee State University, Nashville, TN 37209, USA
}
  
\date{Received date / Accepted date}

\abstract{}
{Previous studies have related surface temperature maps, obtained with the Doppler imaging (DI) technique, of LQ Hya with long-term photometry. Here, we will compare surface magnetic field maps, obtained with the Zeeman Doppler imaging (ZDI) technique, with contemporaneous photometry, with the aim of quantifying the star's magnetic cycle characteristics.}
{We inverted Stokes $IV$ spectropolarimetry, obtained with the HARPSpol and ESPaDOnS instruments, into magnetic field and surface brightness maps using a tomographic inversion code that models high signal-to-noise ratio mean line profiles produced by the least squares deconvolution (LSD) technique. The maps were compared against long-term ground-based photometry acquired with the T3 0.40 m Automatic Photoelectric Telescope (APT) at Fairborn Observatory, which offers a proxy for the spot cycle of the star, as well as with chromospheric \ion{Ca}{ii} H\&K activity derived from the observed spectra.}
{The magnetic field and surface brightness maps reveal similar patterns relative to previous DI and ZDI studies: non-axisymmetric polar magnetic field structure, void of fields at mid-latitudes, and a complex structure in the equatorial regions. There is a weak but clear tendency of the polar structures to be linked with a strong radial field and the equatorial ones with the azimuthal field. We find a polarity reversal in the radial field between 2016 and 2017 that is coincident with a spot minimum seen in the long-term photometry, although the precise relation of chromospheric activity to the spot activity remains complex and unclear. The inverted field strengths cannot be easily related with the observed spottedness, but we find that they are partially connected to the retrieved field complexity.}
{This field topology and the dominance of the poloidal field component, when compared to global magnetoconvection models for rapidly rotating young suns, could be explained by a turbulent dynamo, where differential rotation does not play a major role (so-called $\alpha^2 \Omega$ or $\alpha^2$ dynamos) and axi- and non-axisymmetric modes are excited simultaneously. The complex equatorial magnetic field structure could arise from the twisted (helical) wreaths often seen in these simulations, while the polar feature would be connected to the mostly poloidal non-axisymmetric component that has a smooth spatial structure.}


\maketitle

\section{Introduction}

One of the outstanding features of the 11-year solar activity cycle is the presence of a polarity reversal in the large-scale magnetic field, leading to the 22-year-long magnetic Hale cycle. This magnetic cycle has been one of the key phenomena that has been studies in attempts to explain the operation of the underlying dynamo mechanism \citep[see e.g.][and references therein]{Ossendrijver2003SolarDynamo}. For a comprehensive and general theory of stellar dynamos, a successful dynamo model should not only reproduce the solar cycle, but it should also be applicable to other active stars, whose dynamos operate in different regions of the parameter space. Hence, studying a wide variety of active stars, and their cycles, is crucially important for the understanding of stellar dynamos.

For this reason, there has been a considerable effort to follow the evolution of the magnetic field topology of several active stars using the Zeeman Doppler imaging (ZDI) method \citep{Semel1989ZDI,Brown1991ZDI,Piskunov2002ZDI}. The most well-known star in this regard is \object{$\tau$ Boo}: It has both a short magnetic cycle \citep{Donati2008tauBoo,Fares2009tauBoo,Mengel2016tauBoo} and a 120 d chromospheric cycle \citep{Mengel2016tauBoo,Mittag2017tauBoo}, which have been linked together \citep{Jeffers2018tauBoo}. Another star with observationally linked magnetic and chromospheric cycles is \object{61 Cyg A} \citep{BoroSaikia201661CygA,BoroSaikia201861CygA}. Other stars where magnetic polarity reversals have been reported are \object{HD 190771} \citep{Petit2009HD190771}, \object{HD 78366}, \object{HD 190771}, and \object{$\xi$ Boo A} \citep{Morgenthaler2011PolarityReversals}, and \object{HD 29615} \citep{Waite2015ZDI,Hackman2016ZDI}.

The exact nature of the field topology evolution of stars other than the Sun remains poorly understood; in most cases, it is not clear how the polarity reversals, seen through the limited time windows of the imaging results, relate to the chromospheric or spot activity cycles of the stars. For example, \cite{berdyugina2002} reported a regular 5.2-year flip-flop cycle in LQ Hya from their Doppler imaging (DI) study. Although polarity reversals cannot be detected from DI, such an event implies a drastic and abrupt change in the magnetic field topology. When \cite{Olspert2015LQHya} tried to relate such flip-flop events to the long-term analysis of photometry, revealing the stellar activity cycle, the flip-flops appeared to be completely randomly distributed over the cycle. Relating together the topology changes seen in the ZDI maps is a challenging task as in many cases the cycle periods are long \citep[see e.g.][]{Olspert2018Cycles}, the observational windows are narrow, and the data collected are sparse. Combining long-term photometry and chromospheric activity time series with ZDI and DI inversions seems the most promising way forwards.

Here we present four new ZDI maps of the surface magnetic field of the young rapidly rotating K2V dwarf \object{LQ Hya} (\object{HD 82558}, $V = 7.82$, $B-V = 0.933$ \citep{ESA1997Hipparcos}) to shed light on the evolution of its magnetic field. With a rotation period of 1.60~d \citep{Jetsu1993LQHya}, projected rotational velocity $v\sin i = 26.5 \pm 0.5$~km/s \citep{Donati1999LQHya}, and an age around 50 Myr \citep{Tetzlaff2011YoungStars}, this star can be seen to qualitatively represent the young Sun at the start of its main-sequence evolution. The atmospheric parameters $T_{\rm eff} = 4934 \pm 70$~K, $\log g = 4.50 \pm 0.21$, and $\rm [Fe/H] = -0.14 \pm 0.11$ \citep{Tsantaki2014Parameters} are also close to the solar values, except for the lower effective temperature, which is due to the star's smaller mass. The star is one of the most active solar-type stars to have been studied, showing very high levels of activity both in chromospheric emission, $\log R'_{\rm HK} = -4.03$ \citep{White2007Activity}, and in coronal X-rays, $\log L_{\rm X}/L_{\rm bol} = -3.06$ \citep{Sterzik1997CoolStars}.

Previous ZDI inversions of LQ Hya have been published by \cite{Donati1999LQHya} and \cite{Donati2003LQHya} for epochs between 1991 and 2001. These inversions have revealed average to high strengths of the large-scale magnetic field, reaching about 100 G, and individual field components extending as high as 900 G. The star has also been the subject of extensive DI studies with the aim of mapping the changing spot distribution on its surface \citep{Strassmeier1993LQHya,Rice1998LQHya,berdyugina2002,Kovari2004LQHya,Cole2015LQHya,
FloresSoriano2017LQHya,ColeKodikara2019LQHya}. The picture emerging from these inversions is that the star has exhibited, to a varying extent, both high-latitude and equatorial spots over the years with no detections of mid-latitude temperature anomalies, indicating a possible latitudinal bimodality in their distribution.

Even though there have been detailed studies aiming to relate the DI temperature inversions with the long-term activity evolution of LQ Hya \citep{Cole2015LQHya,Olspert2015LQHya,ColeKodikara2019LQHya}, the same has not been attempted with the magnetic ZDI inversions. In this paper we fill in this gap by comparing our ZDI maps with simultaneous long-term photometry, that traces the spottedness of the stellar surface, and chromospheric activity derived from the \ion{Ca}{ii} H\&K line core emission. As a result we can, for the first time, relate the observed field evolution of LQ Hya with its activity cycle.

\section{Observations}

\subsection{Photometry}

\begin{figure}
\includegraphics[width=\linewidth]{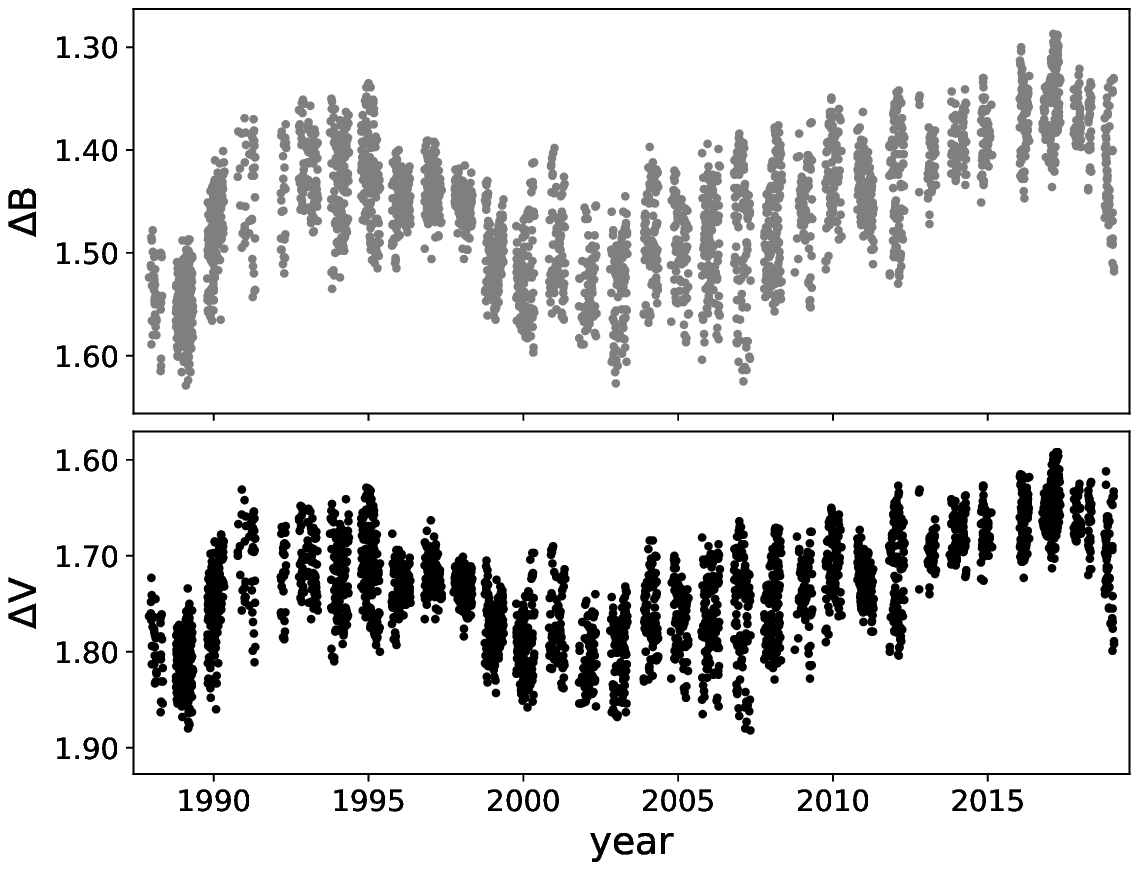}
\caption{Johnson B- and V-band differential photometry of LQ Hya.}
\label{fig-phot}
\end{figure}

\begin{figure}
\includegraphics[width=\linewidth]{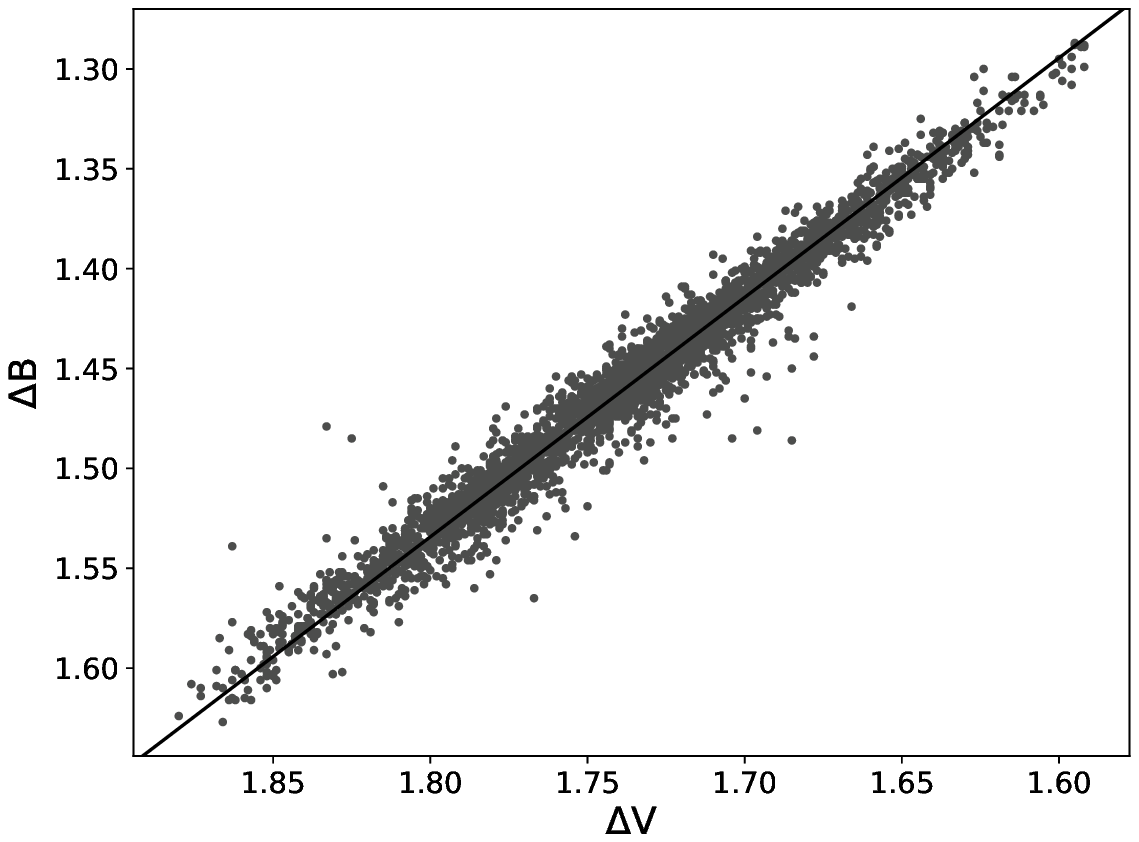}
\caption{B-band vs.~V-band differential photometry of LQ Hya with a line denoting the empirical scaling of Eq.~\ref{eq-phot} between the two bands.}
\label{fig-corr}
\end{figure}

As a proxy for monitoring the spot activity on LQ Hya, we used long-term Johnson B- and V-band differential photometry, which is shown in Fig.~\ref{fig-phot}. These data were acquired using the T3 0.4 m Automatic Photoelectric Telescope (APT) at the Fairborn Observatory in Arizona \citep[see][]{Henry1995APT,Fekel2005Photometry} and span from November 1987 to February 2019. Earlier subsets of the same time series have been studied by \cite{Lehtinen2012LQHya} and \cite{Lehtinen2016Photometry}. All the differential photometry was measured against the comparison star \object{HD 82477}.

To simplify the analysis, we transformed all the photometric data into a single system by scaling the differential B-band photometry onto the differential V-band photometry using the relation
\begin{equation}
\Delta V_B = 0.52 + 0.83 \Delta B.
\label{eq-phot}
\end{equation}
\noindent This empirical scaling relation was determined by a least squares fit to coeval observations in the two photometric bands. It is evident from Fig.~\ref{fig-corr} -- which shows the correlation between the coeval B- and V-band photometry, which has a high linear correlation coefficient, $r=0.98$ -- that a linear transformation between the two bands is sufficient. This also shows that the $B-V$ colour scales linearly with the photometric bands as
\begin{equation}
B-V = 0.20V + \rm cst.
\end{equation}
The two bands can thus be combined together without the loss of information beyond this scaling and may be analysed as a single time series. The relation between the bands shows that the star reddens as it becomes dimmer, which is consistent with the photometric variation being produced by low-temperature spots. We further note that there are no signs of flares in the photometry that could have affected the correlation between the two bands.

\subsection{Spectropolarimetry}

\begin{figure*}
\centering
\begin{tabular}{cccc}
\includegraphics[bb=80 380 535 710,width=4.2cm,clip]{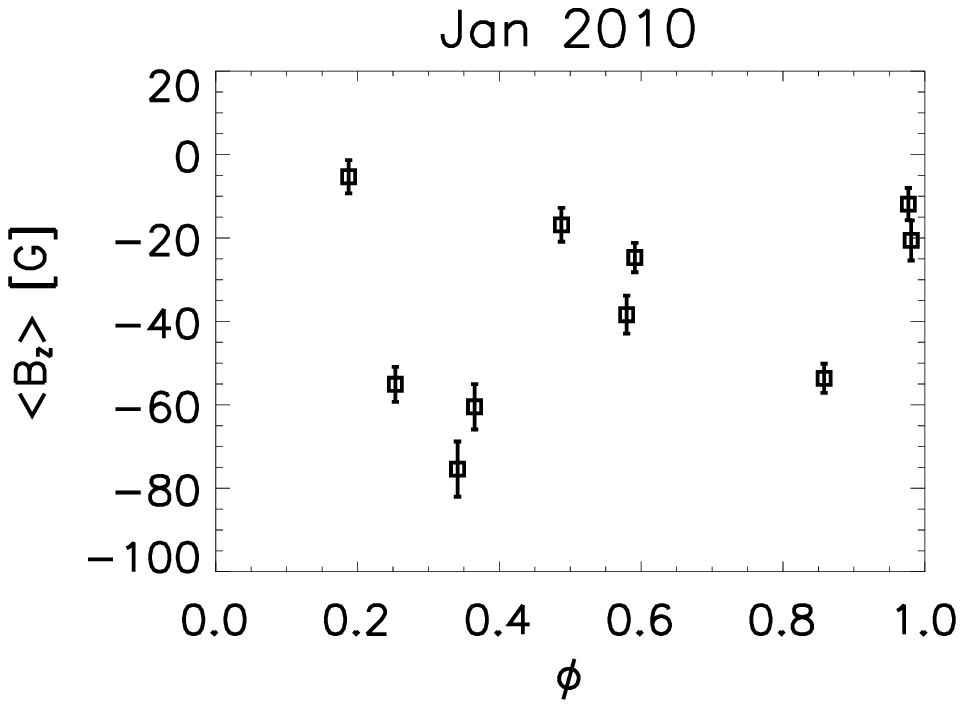} &
\includegraphics[bb=80 380 535 710,width=4.2cm,clip]{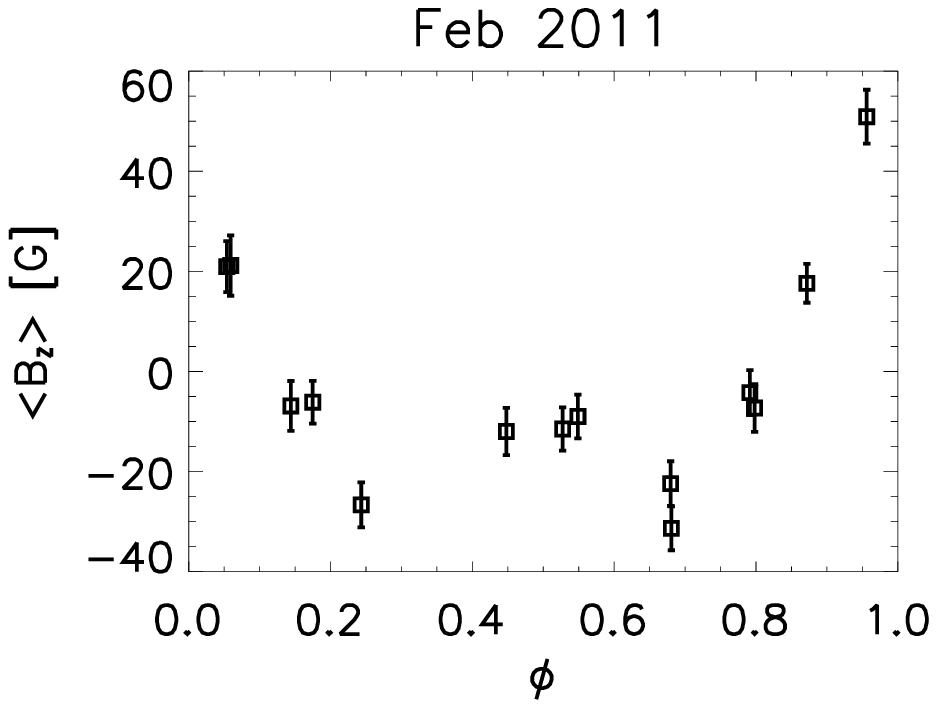} &
\includegraphics[bb=80 380 535 710,width=4.2cm,clip]{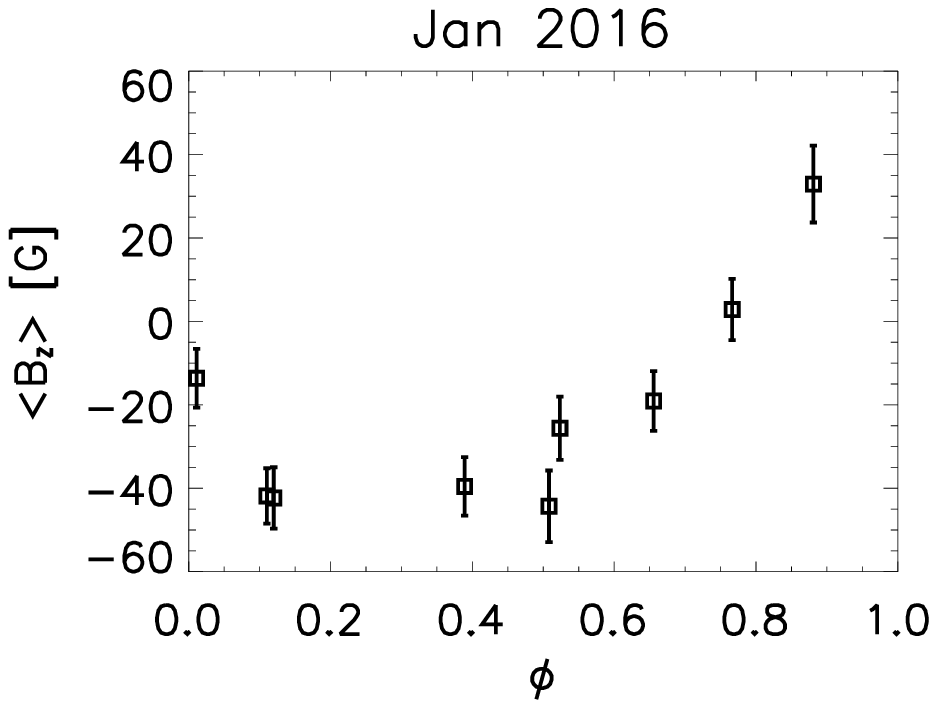} &
\includegraphics[bb=80 380 535 710,width=4.2cm,clip]{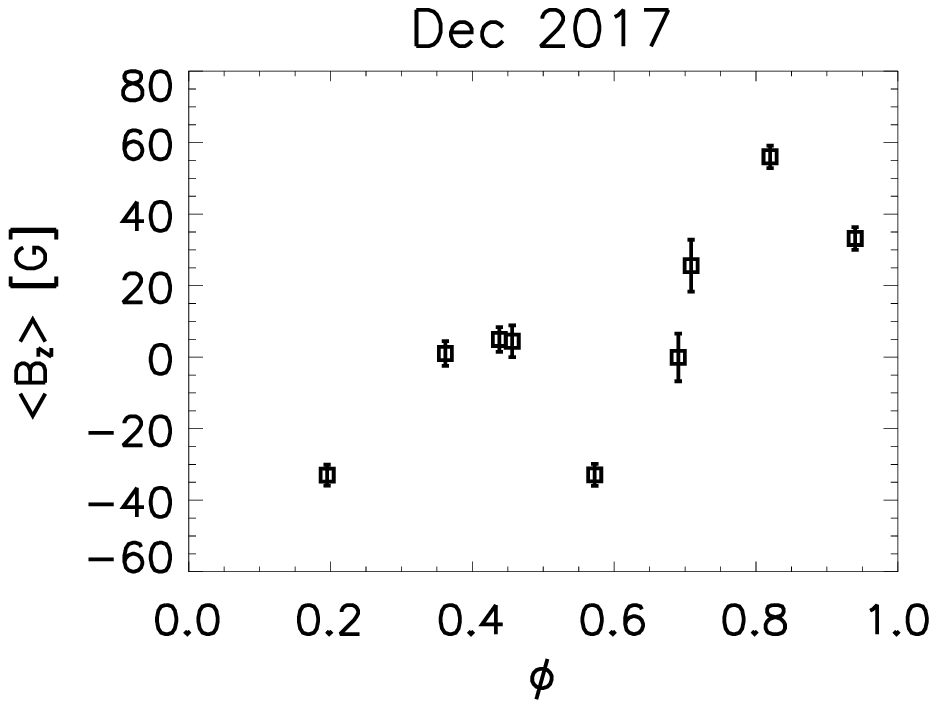}
\end{tabular}
\caption{Mean longitudinal magnetic field $\langle B_z \rangle$ of LQ Hya calculated from the Stokes $V$ LSD profiles as a function of rotation phase $\phi$. The four panels present the $\langle B_z \rangle$ values in each of the observing epochs.}
\label{fig-bz}
\end{figure*}

\begin{figure*}
\centering
\begin{tabular}{cccc}
\includegraphics[width=\linewidth]{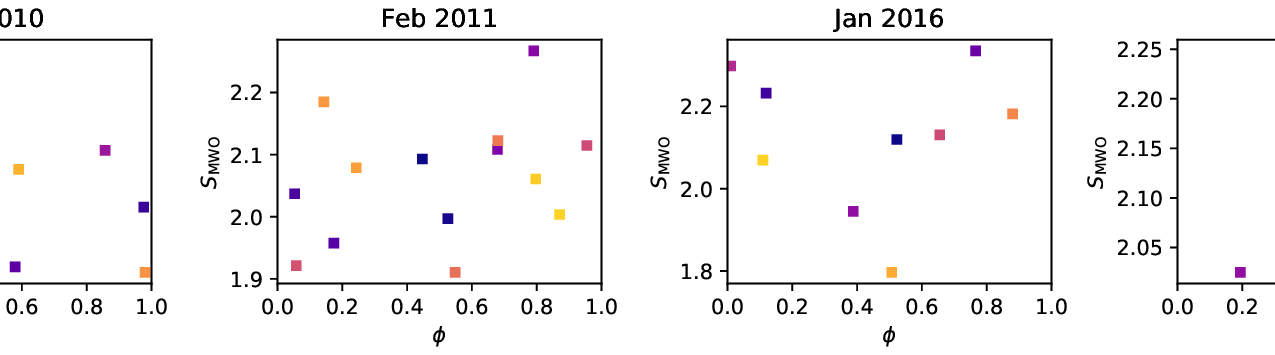}
\end{tabular}
\caption{\ion{Ca}{ii} H\&K activity index $S_{\rm MWO}$ of LQ Hya as a function of rotation phase $\phi$. The four panels present the $S_{\rm MWO}$ values in each of the observing epochs, and the marker colours represent normalised time from the first (blue) to the last (yellow) observation within each observing run.}
\label{fig-chrom}
\end{figure*}

\begin{table}
\centering
\caption{Summary of spectropolarimetric observations}
\begin{tabular}{lcccccc}
\hline
\hline
Epoch & $\langle S/N \rangle_{I}$ & $\langle S/N \rangle_{V}$ & $n_{\rm LSD}$ & $\lambda_0 \ [\AA]$ & $n_\phi$ & $f_\phi$ \\
\hline
Jan 2010 & 153 & 19700 & 7776 & 5094 & 10 & 0.68 \\
Feb 2011 & 139 & 16600 & 7803 & 5171 & 14 & 0.86 \\
Jan 2016 & 219 & 15100 & 7792 & 5291 &  9 & 0.63 \\
Dec 2017 & 168 & 13300 & 7804 & 5148 &  9 & 0.72 \\
\hline
\end{tabular}
\tablefoot{Listed are the mean signal-to-noise ratios $\langle S/N \rangle_{I}$ of the Stokes $I$ spectra and $\langle S/N \rangle_{V}$ of the Stokes $V$ LSD profiles, the number of lines $n_{\rm LSD}$ used for the LSD profiles, the mean wavelengths $\lambda_0$ of the LSD profiles, the number of observed phases $n_\phi$, and the estimated phase coverage $f_\phi$ of the observations. Observations from January 2016 are from ESPaDOnS and the rest from HARPSpol.}
\label{tab-obssum}
\end{table}

\begin{table}
\centering
\caption{Adopted stellar parameters.}
\begin{tabular}{ll}
\hline
\hline
Parameter & Value \\
\hline
Effective temperature & $T_{\rm eff} = 5000 \, \rm K$ \\
(unspotted) & \\
Surface gravity & $\log g = 4.5$ \\
Metallicity & $\rm [M/H] = 0$ \\
Microturbulence & $\xi_t = 1.5 \, \rm km \, s^{-1}$ \\
Rotation period & $P_{\rm rot} = 1.601136 \, \rm d$ \\
Rotational velocity & $v\sin i = 26.5 \, \rm km \, s^{-1}$ \\
Inclination & $i = 65^{\circ}$ \\
\hline
\end{tabular}
\tablefoot{Values from \cite{ColeKodikara2019LQHya}}
\label{tab-param}
\end{table}

\begin{table}
\centering
\caption{Activity levels during the observing epochs}
\begin{tabular}{lcc}
\hline
\hline
Epoch & $S_{\rm MWO}$ & $\log R'_{\rm HK}$ \\
\hline
Jan 2010 \vspace{0.5mm} & $2.025^{+0.237}_{-0.114}$ & $-3.885^{+0.026}_{-0.049}$ \\
Feb 2011 \vspace{0.5mm} & $2.061^{+0.206}_{-0.151}$ & $-3.877^{+0.034}_{-0.042}$ \\
Jan 2016 \vspace{0.5mm} & $2.123^{+0.212}_{-0.327}$ & $-3.865^{+0.073}_{-0.043}$ \\
Dec 2017 \vspace{0.5mm} & $2.136^{+0.113}_{-0.111}$ & $-3.861^{+0.024}_{-0.023}$ \\
\hline
\end{tabular}
\tablefoot{The listed values represent mean activity levels $\langle S_{\rm MWO} \rangle$ and $\log \langle R'_{\rm HK} \rangle$ during the observing epochs and the upper and lower error bars the full variability ranges over the stellar rotation.}
\label{tab-chromsum}
\end{table}

For our ZDI inversions, we obtained high-resolution Stokes $IV$ spectropolarimetry from four separate seasons. Data from three of the epochs (January 2010, February 2011, and December 2017) were observed using the HARPSpol spectropolarimeter located at the ESO 3.6 m telescope in La Silla, Chile, while data from January 2016 were observed using the ESPaDOnS spectropolarimeter at the 3.6 m Canada-France-Hawaii Telescope on Mauna Kea, Hawaii. Both instruments provide a high resolving power at about 110\,000 for HARPSpol and 68\,000 for ESPaDOnS in their spectropolarimetric modes. The HARPSpol data were reduced using the REDUCE package \citep{Piskunov2002Reduction}, while the ESPaDOnS data were automatically reduced by the Upena pipeline, based on the Libre-ESpRIT reduction package \citep{Donati1997Spectropolarimetry}.

A summary of each observing epoch is given in Table \ref{tab-obssum}, and the individual spectropolarimetric observations are listed in Table \ref{tab-obs}. Since the ZDI inversions require a high signal-to-noise ratio ($S/N$) of the polarised line profiles, we used the least-squares deconvolution (LSD) technique to combine a large number of photospheric spectral lines and boost the $S/N$ of the mean line profiles \citep{Donati1997Spectropolarimetry,Kochukhov2010LSD}. The average $S/N$ of the Stokes $I$ spectra at $\lambda=5100 \ \AA$ and the Stokes $V$ LSD profiles, the number of lines used, $n_{\rm LSD}$, and their mean wavelength, $\lambda_0$, are listed in Table \ref{tab-obssum}, while the mean Land\'e $g$-factor of the profiles was $g=1.212$. The line mask used for calculating the LSD profiles was extracted form the VALD database \citep{Piskunov1995VALD,Kupka1999VALD2} for wavelengths $3900-7100 \ \AA$ using stellar parameters listed in Table \ref{tab-param}. We excluded lines with depths less than 5\% from the continuum as well as regions surrounding the strongest lines deviating from the average line profile shape.

The reliability of the surface inversions in both DI and ZDI depends on the rotational phase coverage of the observations over the stellar surface. In Table \ref{tab-obssum}, we list the number of observed phases $n_\phi$ and their calculated phase coverage $f_\phi$ for each of the epochs. Here, $f_\phi$ was estimated assuming that each observation covers a phase range $[\phi-0.05,\phi+0.05]$, following \cite{Kochukhov2013IIPeg}. The calculation of the individual rotation phases $\phi$ is described in Sect.~\ref{sect-zdi}. \cite{Hackman2019HD199178} studied the effect of limited phase coverage in their DI study of \object{HD 199178}. They reported that the surface spot distribution was considerably distorted when the phase coverage was less than 50\% and that some apparent artefacts appeared up to a limit of 70\%. These limits depend, however, on the actual underlying distribution of surface spots versus the phase distribution of the observations. In our case, we have decently high phase coverage in each of our epochs, which should ensure the reliability of our inversions.

We calculated the false alarm probabilities (FAPs) for the magnetic field detection for each of our Stokes $V$ LSD profiles using the reduced $\chi^2$ statistics 
\citep{Donati1992CircularPolarisation,Rosen2013LinearPolarisation}. The values listed in Table \ref{tab-obs} are all extremely low, implying a definite detection of magnetic field from each of our observed spectra. We furthermore determined the mean signed longitudinal magnetic field $\langle B_z \rangle$ from the first moment of the Stokes $V$ profiles \citep{Kochukhov2010LSD}, which are likewise listed in Table \ref{tab-obs} and shown in Fig.~\ref{fig-bz} against the rotation phase $\phi$ defined by Eq.~\ref{eq-epoch}.

\subsection{Chromospheric activity}

Furthermore, we derived the chromospheric \ion{Ca}{ii} H\&K activity levels from each of the observed spectra as the Mount Wilson S index, $S_{\rm MWO}$, and the \ion{Ca}{ii} H\&K emission to bolometric flux ratio, $\log R'_{\rm HK}$. Both HARPSpol and ESPaDOnS have different calibrations determined for deriving emission indices from them \citep{Lovis2011HARPSActivity,AstudilloDefru2017MDwarfActivity,Booth2020ESPaDOnSActivity}. We followed the calibration given by \cite{Lovis2011HARPSActivity} for HARPSpol and \cite{Booth2020ESPaDOnSActivity} for ESPaDOnS. The values of the activity indices are listed in Table \ref{tab-chromsum} as averages and full variability ranges over the stellar rotation and in Table \ref{tab-chrom} for each individual spectrum.

Unlike $\langle B_z \rangle$, the activity level lacks a clear dependence on the stellar rotation, as shown in Fig.~\ref{fig-chrom}. Often close phase pairs show drastically different emission levels between adjacent rotation phases, which points to a fast evolution of chromospheric activity. Colour-coding the data points in Fig.~\ref{fig-chrom} by normalised time from the first to the last observation reveals that typically one or more rotations have passed between the observations, but in December 2017 there is even evidence for fast activity evolution within a single night (around $\phi = 0.4$ and $\phi = 0.7$). The seasonal variation of the activity level is included in the top panel of Fig.~\ref{fig-cps} along the photometric mean brightness variation and is discussed in the following section.

\section{Time series analysis of photometry}

\begin{figure}
\includegraphics[width=\linewidth]{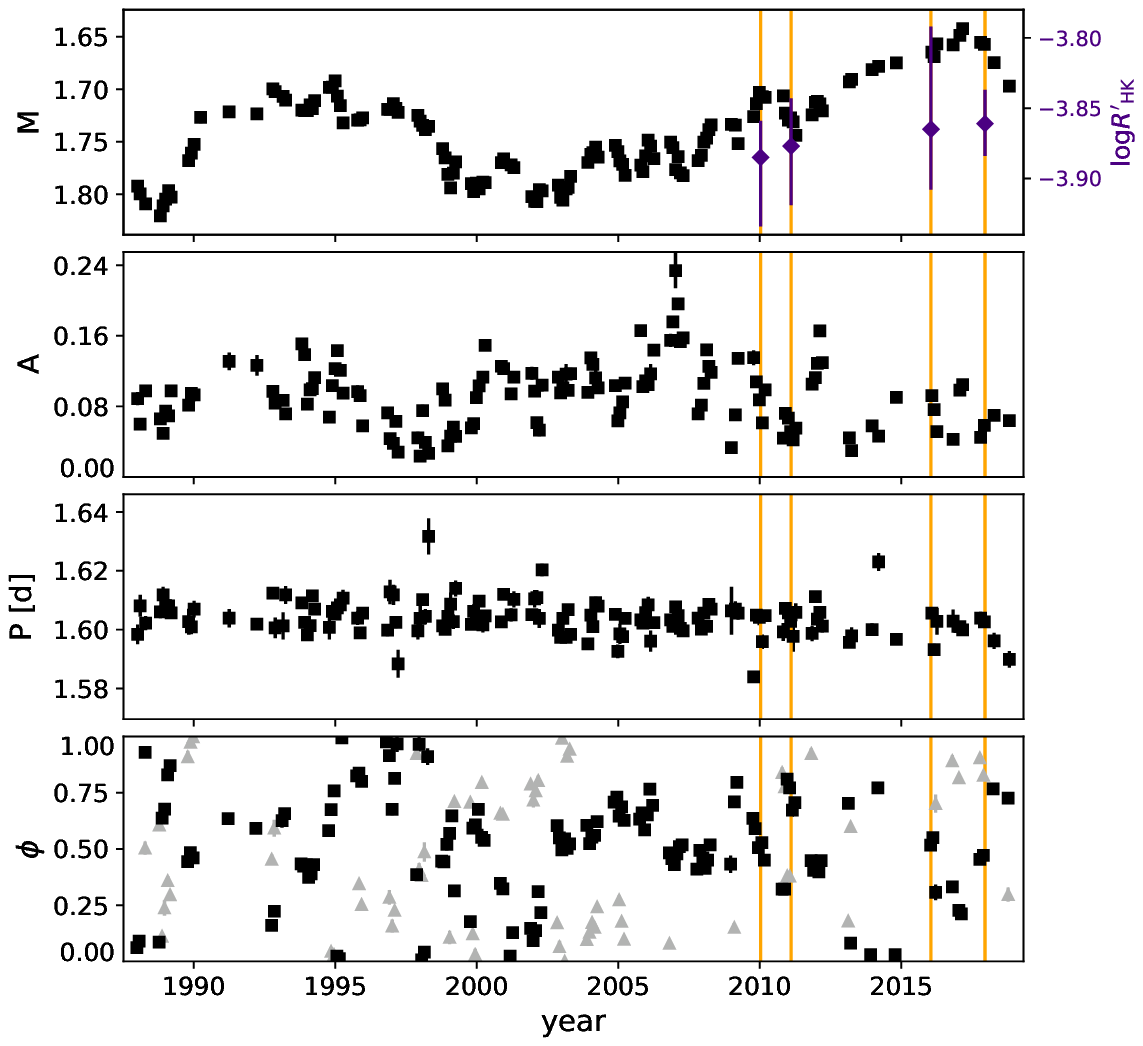}
\caption{Light curve mean magnitude $M$, amplitude $A$, period $P$, and phases of the light curve minima $\phi_{\rm min}$ for LQ Hya. Both $M$ and $A$ are in the scale of $\Delta V$. The minimum phases are shown as black squares for the primary light curve minima and grey triangles for the secondary minima. The vertical orange lines denote the epochs of the ZDI maps. The average chromospheric activity levels $\log R'_{\rm HK}$, corresponding to the ZDI maps, are shown in the top panel as violet diamonds with error bars indicating the variability ranges over the stellar rotation.}
\label{fig-cps}
\end{figure}

The photometric time series was analysed using the Continuous Period Search method \citep{Lehtinen2011CPS}, which searches for low-order hamonic fits to short subsets of the light curve data, chosen using a sliding window. Here we used a window with the length of $\Delta t_{\rm max} = 30$ d and concentrated on only a mutually independent sub-selection of non-overlapping fits. Moreover, all fits where the distribution of the fit residuals or the bootstrap distributions of any of the light curve model parameters were tested to be non-Gaussian using the Kolmogorov-Smirnov test at a significance level of 0.01 were rejected as unreliable \citep{Lehtinen2011CPS}. Examples of individual light curve fits for the photometric time series of LQ Hya can be seen in Fig.~2 of \cite{Lehtinen2012LQHya} up to the year 2011. The results of our time series analysis are presented graphically in Fig. \ref{fig-cps}. Here $M$ denotes the mean level of each light curve fit in the differential V-band, $A$ the full V-band light curve amplitude, and $P$ the light curve period, corresponding to the stellar rotation.

We calculated the weighted mean photometric period of LQ Hya to be $P_{\rm phot} = 1.60405 \pm 0.00040$ d and the relative $\pm 3\sigma$ range of period variability as $Z = 6\Delta P_{\rm phot}/P_{\rm phot} = 0.017$, where $\Delta P_{\rm phot}$ denotes the $1\sigma$ range of the period measurements in the light curve fits \citep{Jetsu1993LQHya}. This quantity gives an approximation of the relative differential rotation coefficient $k = \Delta\Omega/\Omega$ and so denotes nearly solid body rotation, $k \approx 0.017$. Such a low differential rotation is well in line with previous estimates from photometry \citep{Jetsu1993LQHya,berdyugina2002,Messina2003DifferentialRotation,You2007LQHya,Lehtinen2012LQHya,Lehtinen2016Photometry} as well as DI/ZDI studies \citep{Donati2003DifferentialRotation,Kovari2004LQHya}.

A period search of the light curve mean magnitudes $M$ indicates a spot cycle of the length $P_{\rm cyc} = 20.5$ yr. This estimate was obtained following the method description of \cite{Horne1986Period} and was done in an identical way to the cycle search performed by \cite{Lehtinen2016Photometry}. The activity cycle estimates for LQ Hya have not yet converged. Earlier reports indicated a shorter cycle length increasing over time. \cite{Jetsu1993LQHya} found a cycle of 6.24 yr between 1984 and 1992 and \cite{Olah2009Cycles} found that this cycle had increased in length from 7 yr to 12.4 yr by the end of their dataset. Using more recent data, extending to 2014, \cite{Lehtinen2016Photometry} reported longer cycle period estimates ranging from 14.5 yr to 18.0 yr, which approach our present period estimate. There was indeed an extended increase in the mean brightness of LQ Hya from about 2002 to 2016 that, together with a decreasing light curve amplitude, can be attributed to a diminishing spot coverage \citep{ColeKodikara2019LQHya}.

Between 2016 and 2017, the mean brightness finally started to decrease again, suggesting a resurgence of spot activity. Interpreting this brightness maximum as a minimum of spot activity is justified by the observation that the brightness variation of medium to very active stars, like LQ Hya, is spot dominated \citep{Lockwood2007PhotometricVariation,Radick2018PhotometricVariation}. This means that their year-to-year average brightness anticorrelates with the variability of their chromospheric activity, preceeding the brightness maxima in their long-term light curves corresponding to periods of low spottedness. This is supported by \cite{Cao2014LQHya}, who observed that the chromospheric activity of LQ Hya decreased from 2006 to 2012 during the extended phase of increasing brightness. This corresponds to an anticorrelation between activity and brightness. On the other hand, our own measurements of $\log R'_{\rm HK}$ indicate a weak increase in activity from 2010 to 2017, leading to the brightness maximum (see Fig.~\ref{fig-cps}). As such, the precise relation between the spot and chromospheric activity of LQ Hya, as well as the whole nature of its chrompspheric cycle, still remains unclear. The brightness maximum between 2016 and 2017 should thus more accurately be called a spot minimum.

In Fig. \ref{fig-cps} we also show the phases of the light curve minima, $\phi_{\rm min}$, which track the rotational phases of the major spot features on the star. These phases were calculated using the active longitude period $P_{\rm al} = 1.603733$ d, determined by \cite{Lehtinen2016Photometry} as the phasing period that produces the maximal phase concentration of the light curve minima. This period tracks the rotation of a long-lived coherent spot forming structure, that is to say, an active longitude, seen on the star for several years from roughly 2003 to 2008 and displaying modest migration during that time. Both before and after this time period it has not been possible to identify long lasting preferential longitudes of spot formation. The current era of no long-lived active longitudes coincides with the epochs of our ZDI inversions.

\section{Zeeman Doppler imaging}
\label{sect-zdi}

We inverted the Stokes $IV$ profiles into magnetic field and surface brightness maps using the inversLSD code developed by \cite{Kochukhov2014ZDI}. The inversion treats the magnetic field components as spherical harmonic expansions, which allows straightforward calculation of magnetic energy spectra from the inverted maps as well as determining the energy fraction stored in the toroidal and poloidal field components. In this study we restricted the angular degrees of the expansion to $\ell_{\rm max}=20$ and found that for all of our epochs the highest angular orders $\ell=20$ contained less than 1\% of the total magnetic energy of the inverted field. It should be noted, however, that in each inversion most of the magnetic field energy was contained in the low orders ($\ell \le 6$).

We performed the inversion of both the magnetic field and surface brightness maps simultaneously so that the brightness inhomogenities on the star were properly taken into account in the inversion of the magnetic field. This is a necessary procedure to mitigate the effect of reduced polarisation signal from darker surface features, which could otherwise be interpreted to correspond to weaker field strength \citep{Rosen2012ZDITemp}.

The regularisation of the inversion solutions was done as described by \cite{Rosen2016ZDI}. For the field solution we suppressed the high-order modes while for the brightness solution we used Tikhonov regularisation. The strength of regularisation was determined so that the mean deviation between the profile fits and the LSD profiles stayed consistent with the noise level of the profiles. This is crucial for deriving consistent magnetic field strengths from the inversions since the amplitude of the inverted field is sensitive to the strength of the chosen field regularisation. The inversions furthermore included an additional penalty function for the surface brightness, which ensures that the inverted brightness values do not stray too far from their set mean value at 1 \citep[see][]{Hackman2016ZDI}.

The stellar parameters adopted for our inversions are listed in Table \ref{tab-param} and are the same as those used in the DI study of LQ Hya by \cite{ColeKodikara2019LQHya}. Furthermore, we adopted exactly the same ephemeris, originally from \cite{Jetsu1993LQHya}, for calculating the rotational phases of the spectropolarimetric data as in these previous studies:
\begin{equation}
\rm HJD_{\phi=0} = 2\,445\,274.22 + 1.601136 \times E.
\label{eq-epoch}
\end{equation}
The rotation period $P_{\rm rot} = 1.601136$~d used in this ephemeris is slightly different from the rotation period $P_{\rm phot} = 1.60405$~d determined from the current photometry. Within the duration of our individual observing runs this difference accumulates only minimal phase shifts for the observations and mostly results in a different definition for the zero longitude. We chose to use the old ephemeris specifically to guarantee an identical coordinate frame for our ZDI maps with the temperature maps published by \cite{ColeKodikara2019LQHya}, allowing a direct comparison between the two studies. This choice was validated by calculating test inversions using an ephemeris based on $P_{\rm phot}$. In this test we concluded that the differences between inversions using the two periods are, indeed, primarily phase shifts related to choosing a different coordinate frame. Remaining differences are attributable to minor reshuffling of the observing phases in cases where there may have been fast magnetic field evolution on the star. In such situations, the inversions produce in any case less reliable solutions than during epochs of more stable field.

The line profile fits to our Stokes $IV$ LSD profiles resulting from the inversions are shown in Fig. \ref{fig-prof}. In general the fits fall close to the noise level of the profiles. In one case of close phase pairs separated by several rotations of the star, there has been a noticeable evolution in the Stokes $V$ profile. The phase pair in question is $\phi=0.976$ and $\phi=0.980$ in January 2010, which is separated by five stellar rotations between the observations. In this case there was some field evolution between the two observing epochs. Such rapid changes may pose uncertainties to the inversion solutions in the form of smaller details in the derived maps.

\section{Results}

\begin{table*}
\caption{Summary of the magnetic field inversions.}
\centering
\begin{tabular}{lcccccccc}
\hline \hline
Epoch & $\langle|B|\rangle$ & $|B|_{\max}$  & Poloidal & Toroidal & Axisym.&
Non-axisym. & $\ell=1$ & Peak $\ell$ \\
      & [G] & [G] & [\%] & [\%] & [\%] & [\%] & [\%] \\ 
\hline
Jan 2010 & 256 & 1182 & 69 & 31 & 52 & 48 & 23 & 1 \\
Feb 2011 & 169 &  526 & 33 & 67 & 55 & 45 & 32 & 1 \\
Jan 2016 & 163 &  609 & 75 & 25 & 47 & 53 & 26 & 1 \\
Dec 2017 & 157 &  605 & 79 & 21 & 17 & 83 & 19 & 2 \\
\hline
\end{tabular}
\tablefoot{The given percentages denote the fraction of magnetic energy stored in the poloidal, toroidal, axisymmetric, and non-axisymmetric magnetic field components and the $\ell=1$ spherical harmonic modes. The peak $\ell$ denotes the angular order containing most of the magnetic energy.}
\label{tab-magn}
\end{table*}

\begin{figure*}
\centering
\begin{tabular}{c}
\includegraphics[bb=255 40 400 812,width=3.5cm,clip,angle=90]{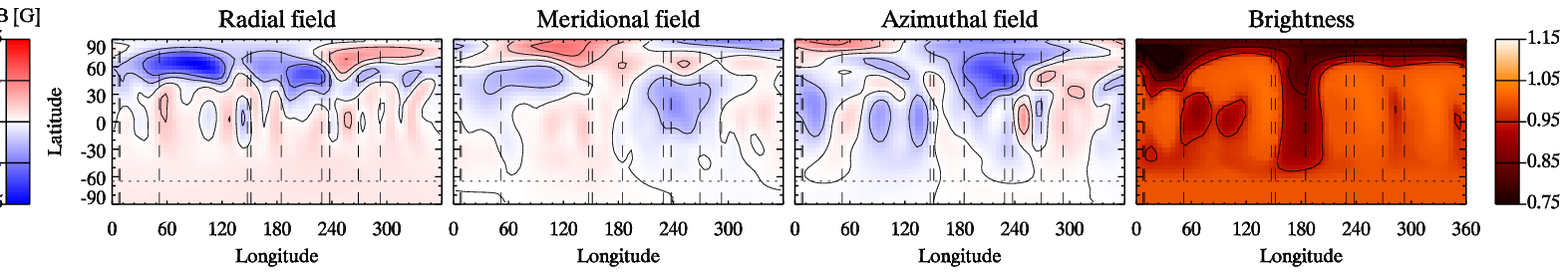}
\vspace{2mm} \\
\includegraphics[bb=255 40 400 812,width=3.5cm,clip,angle=90]{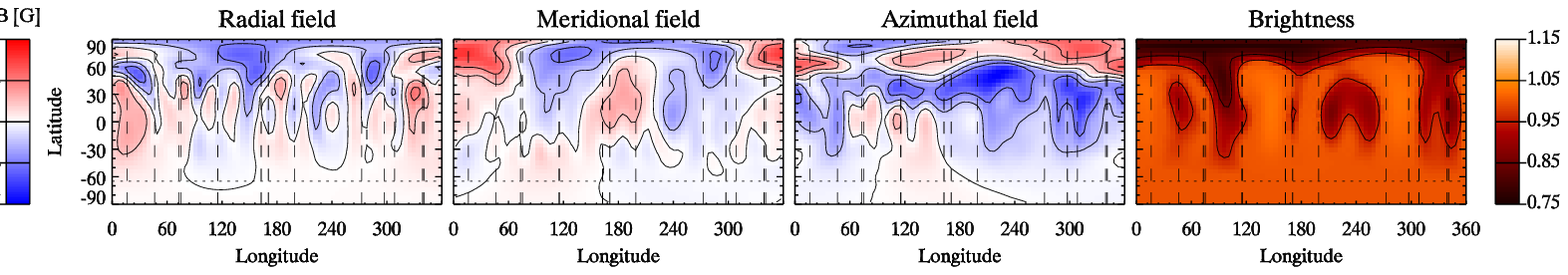}
\vspace{2mm} \\
\includegraphics[bb=255 40 400 812,width=3.5cm,clip,angle=90]{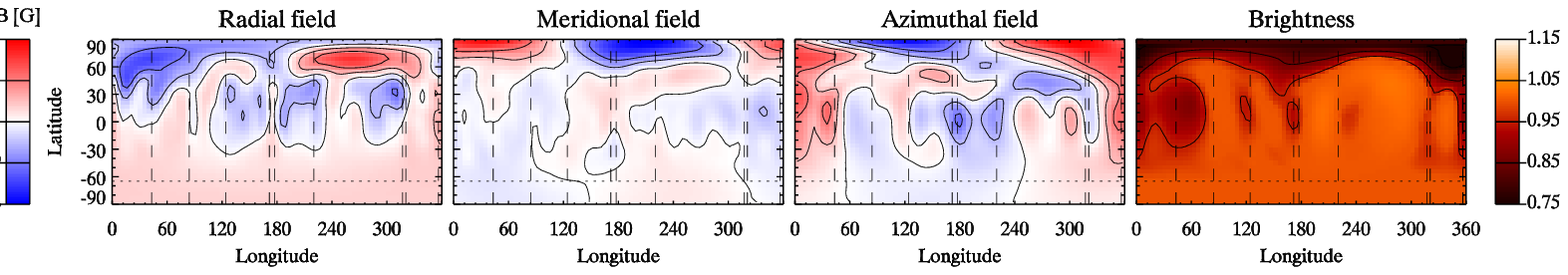}
\vspace{2mm} \\
\includegraphics[bb=255 40 400 812,width=3.5cm,clip,angle=90]{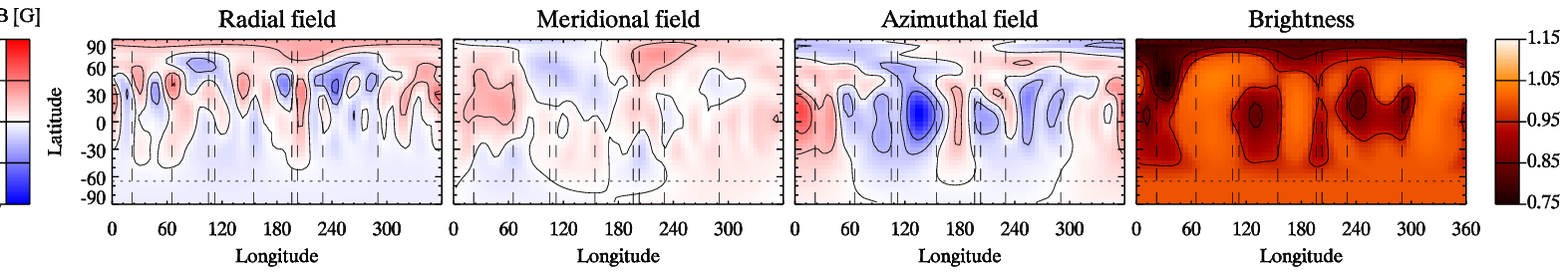}
\vspace{2mm} \\
\end{tabular}
\caption{\textit{Top to bottom:} ZDI maps of LQ Hya in equirectangular projection from the epochs January 2010, February 2011, January 2016, and December 2017. In each epoch separate maps are shown for the radial, meridional, and azimuthal magnetic field components as well as the surface brightness distribution. The vertical lines in the maps denote the rotational phases of the individual observation and the horizontal line the lower limit of visibility due to the stellar inclination.}
\label{fig-map}
\end{figure*}

\begin{table*}
\caption{Correlations between the field component amplitudes and surface brightness.}
\centering
\begin{tabular}{lccccccccc}
\hline \hline
& \multicolumn{3}{l}{Global} & \multicolumn{3}{l}{Lat. $\ge 45^\circ$} & \multicolumn{3}{l}{Lat. $< 45^\circ$} \\
Epoch & $r(|B_r|,b)$ & $r(|B_m|,b)$ & $r(|B_a|,b)$ & $r(|B_r|,b)$ & $r(|B_m|,b)$ & $r(|B_a|,b)$ & $r(|B_r|,b)$ & $r(|B_m|,b)$ & $r(|B_a|,b)$ \\
\hline
Jan 2010 & $-0.22$ & $\hphantom{-}0.04$ & $-0.07$ & $-0.04$ & $\hphantom{-}0.03$ & $-0.01$ & $-0.06$ & $\hphantom{-}0.25$ & $\hphantom{-}0.07$ \\
Feb 2011 & $-0.22$ & $-0.34$ & $-0.17$ & $-0.08$ & $-0.26$ & $\hphantom{-}0.19$ & $-0.08$ & $-0.26$ & $-0.26$ \\
Jan 2016 & $-0.21$ & $-0.45$ & $-0.59$ & $-0.04$ & $-0.54$ & $-0.66$ & $-0.03$ & $\hphantom{-}0.04$ & $-0.47$ \\
Dec 2017 & $-0.10$ & $-0.24$ & $-0.41$ & $-0.06$ & $-0.25$ & $\hphantom{-}0.05$ & $-0.07$ & $-0.07$ & $-0.56$ \\
\hline
\end{tabular}
\tablefoot{Correlations are given for the radial $|B_r|$, meridional $|B_m|$, and azimuthal $|B_a|$ magnetic field amplitudes both over the full stellar surface and separately for latitude bands higher and lower than $45^\circ$.}
\label{tab-corr}
\end{table*}

\subsection{Magnetic topology}

The ZDI maps are presented in Fig. \ref{fig-map} and a summary of their field topology is given in Table \ref{tab-magn}. The inversions reveal a predominantly poloidal magnetic field configuration with mostly equal fractions of field energy in the axisymmetric and non-axisymmetric harmonic components. The division into the axisymmetric and non-axisymmetric parts was done following \cite{Fares2009tauBoo} so that the spherical harmonic modes with $m < \ell/2$ were considered axisymmetric and the remaining ones non-axisymmetric. The inverted fields have most of their energy in the low angular orders, as can be seen from the peak $\ell$ values listed in Table \ref{tab-magn} and the magnetic energy spectra in Fig. \ref{fig-el}. However, the maps still reveal a complex field configuration and in none of the cases the $\ell=1$ mode dominates the inverted field. One notable small-scale feature in the magnetic maps are the low-latitude bands of small-scale fluctuations in the radial field.

The fact that most of the maps are neither strongly axisymmetric or non-axisymmetric may be connected to the intermittent nature of the active longitudes seen in the photometric analysis of LQ Hya \citep{Olspert2015LQHya,Lehtinen2016Photometry} and may be a sign of competing axisymmetric and non-axisymmetric dynamo modes. In the December 2017 map, the field turned strongly non-axisymmetric, and it remains to be seen whether this change will signify a return to a state of longer lived active longitudes on the star.

Between the January 2016 and December 2017 maps there was a complex shift in the magnetic field topology. Firstly, the radial field $B_r$, which had shown predominantly negative polarity around the visible pole and positive polarity on the opposite hemisphere, reversed its sign between these epochs. This polarity reversal can also be seen in the mean longitudinal field, $\langle B_z \rangle$ (see Fig. \ref{fig-bz}), which changes from mainly negative to mainly positive in the last epoch. Secondly, both the meridional and azimuthal fields $B_m$ and $B_a$, which had previously been strongest at the high latitudes, lack a strong polar component in the last map. Instead, the azimuthal field is dominated by a complex equatorial structure in this map.

What makes these topological changes particularly interesting is that they occurred exactly at the same time as the minimum in the spot activity. The epochs of the ZDI maps are marked in Fig. \ref{fig-cps} with vertical lines over the photometric results. Here it is evident that the January 2016 and December 2017 maps flank roughly symmetrically the epoch of minimum spottedness. The polarity reversal is thus highly likely connected with the minimum of the spot cycle.

Comparing our maps with the ZDI inversions of \cite{Donati1999LQHya} and \cite{Donati2003LQHya} is in many cases difficult since several of their maps have poor phase coverage. But where their maps have good phase coverage, they show predominantly negative polarity of the radial field around the visible pole. The epochs of these maps place them mostly after the previous spot minimum that seems to have occurred some time between 1993 and 1995. As such, they would be consistent with the interpretation that the radial field of LQ Hya retained a constant polarity throughout the whole 20-year spot cycle, only switching over during the consequent minimum.

In Table \ref{tab-corr}, we present the correlations between the amplitudes of each of the field components with the surface brightness maps. These correlations have been calculated both globally and separately for latitudes above and below $45^\circ$. We can see that in most cases there is a lack of correlation between the field and the spots, which is a well-known feature of ZDI inversions \citep[see e.g.][]{Kochukhov2013IIPeg,Waite2015ZDI,Hackman2016ZDI}. When we do see stronger correlation, it is anticorrelation between the field and surface brightness, denoting magnetic field that concentrates in the dark spots. In these cases the strongest anticorrelation can be found with the azimuthal field. In the December 2017 map, the azimuthal field follows the low-latitude spots so clearly that the strongest field patches at longitudes $0^\circ$ and $140^\circ$ can be visually identified with the corresponding spots. We note, however, that even in the cases of the strongest anticorrelation, it remains weak, pointing to a general decorrelation between the large-scale field and the surface brightness. A discrepancy between high and low latitudes may indicate that the polar and low-latitude spots are truly distinct from each other and have their origin in the different components of the dynamo field.

\subsection{Comparison with temperature maps}

In \cite{ColeKodikara2019LQHya} two nearly coeval (their December 2009 and December 2017) and two nearly simultaneous (their December 2010 and December 2015) DI temperature maps were presented. Their 2009 and 2010 data were obtained with the SOFIN instrument, and the 2015 and 2017 data with FIES at the Nordic Optical Telescope. 

In December 2009, \cite{ColeKodikara2019LQHya} recovered strong high-latitude spots around the phase $\phi=0.75$, forming an asymmetric polar cap. As their phase axis is mirrored with respect to our longitude axis, this would corresponds to longitude 90$^{\circ}$ in our scale. We see a good correspondence of this feature with the strongest location of the radial field in our January 2010 ZDI map. Due to the poor phase coverage, the equatorial structures recovered in the DI maps were judged to be artefacts, but in our ZDI brightness maps they appear again and seem to have a weak preferential association to the azimuthal magnetic field.

In December 2010, the phase coverage of \cite{ColeKodikara2019LQHya} was poor again and perhaps for that reason they saw no high-latitude activity. The correspondence of their DI map with the magnetic field configuration in our ZDI map of February 2011 is therefore poor. On the other hand, it is interesting to note that they detected strong line-profile variability during this season. This is an indication of rapid changes in the spot configuration. The DI and ZDI maps were, furthermore, derived from data observed almost 1.5 months apart, which might also contribute to the mismatch.

In December 2015, the phase coverage of \cite{ColeKodikara2019LQHya} was decent, and they recovered strong activity at high latitudes, while the equatorial spots appeared as artefacts in the DI map. Due to the large gap in phases around $\phi=0.5$, it was difficult to determine the longitudinal extent of the spot structure. In comparison to our January 2016 ZDI map, the best coincidence of the temperature minima in the DI and ZDI maps occurred for the meridional field component.

In December 2017, the best quality observing season of \cite{ColeKodikara2019LQHya}, only high-latitude spot activity was recovered. The maximal spot activity concentrated near phase $\phi=0.4$ in their DI map. This, again, matches with the strongest radial field component in our December 2017 ZDI map, but not with the equatorial structures seen in the brightness map and azimuthal component of the field.

In summary, the comparison of the DI maps of \cite{ColeKodikara2019LQHya} and our ZDI maps indicate the following picture: The high-latitude spot structures that are the most pronounced features in the DI maps are related to the radial/meridional magnetic field components in the ZDI maps. The deviation between the DI and ZDI maps is largest in the equatorial regions, where detectable magnetic fields are mostly related to the azimuthal component of the magnetic field. The equatorial structures are prominent in the ZDI brightness maps, but the corresponding temperature features are either uncertain or absent from the DI maps.

\subsection{Field strength and complexity}

Although the high mean and maximum field strengths $\langle|B|\rangle$ and $|B|_{\rm max}$ of the ZDI maps, as listed in Table \ref{tab-magn}, are of similar magnitude as what was seen by \cite{Donati1999LQHya} and \cite{Donati2003LQHya}, the variability shown by $|B|_{\rm max}$ is considerable. It seems unlikely that such high actual variability could happen at the length scales of the inverted field within the short time intervals between the observed epochs.

One possible explanation for the apparent peak field variability is that it does not represent actual field strength variability but rather results from us retrieving a variable fraction of the total surface field in the different maps. In this case it is reasonable to assume that there would be a relation between field complexity and the peak field strength as the more complex inverted field structures would capture more of the small to intermediate scale field energy.

\begin{figure}
\includegraphics[width=\linewidth]{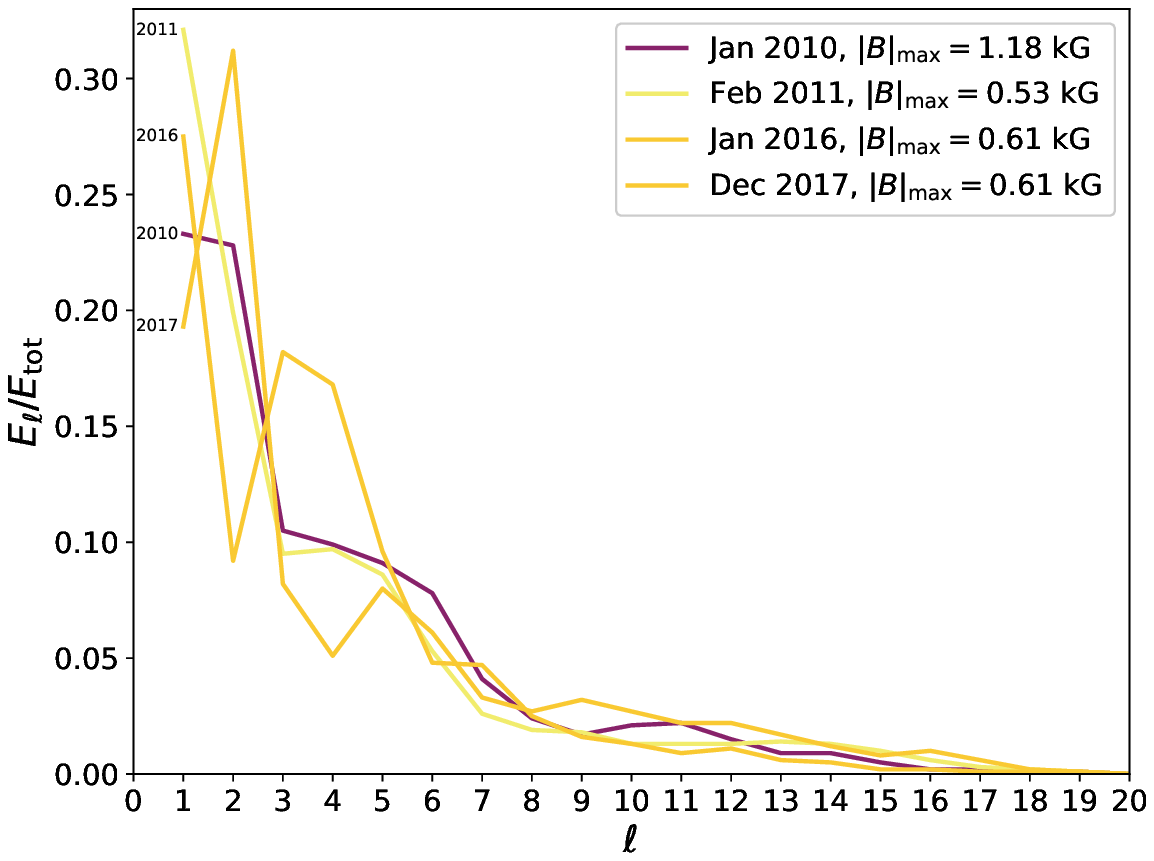}
\caption{Magnetic energy spectra of the ZDI maps as a function of the angular order $\ell$, normalised by the total magnetic energy $E_{\rm tot}$. The line colours indicate the maximum field strength $|B|_{\rm max}$ in each map, as given in the legend, darker colours denoting stronger fields.}
\label{fig-el}
\end{figure}

This hypothesis is investigated in Fig. \ref{fig-el} by comparing the energy spectra over the angular order $\ell$ of the four ZDI maps with their peak field strengths. In the three maps from February 2011 to December 2017 there is a possible, although weak, connection between the field complexity and $|B|_{\rm max}$. The stronger peak fields in the January 2016 and December 2017 maps correspond to significant peaks of magnetic energy in the energy spectra at orders $\ell=2$ and $\ell=3$. Similar peaks of magnetic energy at higher $\ell$ orders are absent in the weaker field map of February 2011. This suggests that the apparent variability of the peak field is at least partly governed by the field complexity of the inversion solution.

On the other hand, the January 2010 map breaks this pattern by having most of its field energy in the low orders $\ell=1$ and $\ell=2$ despite also having the highest $|B|_{\rm max}$ of all the maps. Here we have to conclude that we are dealing with a genuinely stronger underlying field. This interpretation fits well with the fact that this epoch is also the one furthest in time from the spot minimum in 2016--2017 and also has higher amplitude spot modulation in the light curve than the rest of the epochs. In the test inversions, calculated using the photometric period $P_{\rm phot}$, the peak field strength of the January 2010 map was somewhat lower than the peak field in the corresponding inversion calculated using $P_{\rm rot}$ (1.04 kG as opposed to 1.18 kG), but was still significantly stronger than in the later maps. This difference may be symptomatic of the fast changing field configuration during this epoch (see Sect. \ref{sect-zdi}), introducing uncertainty in the inversion solution.

Moreover, if we look at the mean field stregths $\langle|B|\rangle$, we see that in January 2010 they were elevated from the more modest values they had during the later epochs. It seems thus that the high peak field strength during January 2010 had more to do with a higher physical field strength on the star while the variability seen during the later epochs is, to a larger extent, related to the level of detail that was resolvable in the inversions. Thus, we do not find the inverted field strength in the later epochs to be indicative of intrinsic field strength variations.

\section{Conclusions}

The four new ZDI inversions of LQ Hya provide a number of interesting results concerning the magnetic field evolution and spot structure. The central observation from these maps is a marked change in the magnetic field topology between 2016 and 2017, occurring at the same time as a spot minimum. During this time we can see both a polarity reversal of the radial field and a shift of the strongest azimuthal field from high latitudes to the equatorial region. This change is a fairly complex one and cannot currently be fully understood. If our assumption is correct that at least the polarity reversal of the radial field is connected with the spot minimum seen in the photometry, we would expect not to see another similar change in quick succession due to the extended length of the star's activity cycle. Comparing our maps with those of \cite{Donati1999LQHya} and \cite{Donati2003LQHya} suggests that this may have been the case for the recently ended cycle, which seems to have preserved the radial field polarity all the way from its beginning.

The magnetic field strengths retrieved by the inversions vary significantly between the maps and cannot easily be related to the level of spottedness or the chromospheric activity. We are left to conclude that this is likely partly an effect of varying complexity of the inverted field that leads to a different fraction of the true physical field strength being retrieved in the maps. Regardless, the field strengths in our inversions are consistent with those observed by \cite{Donati1999LQHya} and \cite{Donati2003LQHya}.

Comparison between our ZDI surface brightness maps and the simultaneous or nearly simultaneous DI temperature maps of \cite{ColeKodikara2019LQHya} builds a tentatively optimistic picture of the reliability of the equatorial spots seen in the inversions. These features have not always been taken to represent real physical spots on the star since with poorer quality datasets they can often be explained as artefacts of the inversion procedure. In our case we often see equatorial spots appearing in both the DI and ZDI maps, and in our ZDI inversions we also see a weak connection between the azimuthal magnetic field and the equatorial spots, which both raise the likelihood that these features represent real physical features on the stellar surface. If this is the case, we are left to conclude that the spot structure of LQ Hya is bimodal in latitude, consisting of strong high-latitude to polar spots and a weaker band of equatorial spots. The polar spots are mostly connected with the radial and meridional field while the equatorial spots seem to have an association with the azimuthal field. This may indicate a difference in the originating processes between the spots.

The magnetic field structures retrieved in the ZDI maps match reasonably well the recent global magnetoconvection models for stars. For example, in the works of \cite{Gastine2012Dynamo} and \cite{Viviani2018Axisymmetry}, the radial and meridional components of the magnetic field are seen to concentrate mostly near the poles in a non-axisymmetric configuration, while `wreaths' of azimuthal magnetic field are generated near the equatorial regions. Especially in the models of \cite{Gastine2012Dynamo} the equatorial azimuthal field was exhibiting a high degree of twisting, which would be seen as a complicated multi-polar structure of the surface magnetic field if these fields emerged to the surface to form spots. Such solutions are mainly retrieved for models where the rotational influence is moderate. Polarity reversals, however, are very rare, while azimuthal dynamo waves occur nearly always, but are not so pronounced in LQ Hya according to the observations \citep{Olspert2015LQHya,Lehtinen2016Photometry}. One must bear in mind, however, that these models are still far removed from the parameter ranges of real stars.

In more general terms of dynamo theory, the poloidal magnetic field component being almost always dominant to the toroidal one, gives a clear indication that the dynamo in this star is not of $\alpha \Omega$ type as in the Sun. Whereas differential rotation is known to play a crucial role for the solar dynamo, observational indications for surface differential rotation in LQ Hya are weak. This also agrees with the results from global magnetoconvection models \citep[e.g.][]{Viviani2018Axisymmetry}, where the absolute differential rotation becomes small in the rapid rotation regime. In this case, the generation of both poloidal and toroidal magnetic fields occurs by the action of turbulent-convection-produced $\alpha$ effect (either a $\alpha^2 \Omega$ or $\alpha^2$ dynamo).

\begin{acknowledgements}
MJK and JJL acknowledge the Academy of Finland `ReSoLVE'. Centre of Excellence (project number 307411) and the Max Planck Research Group `SOLSTAR' funding. TH acknowledges the financial support from the Academy of Finland for the project SOLSTICE (decision No. 324161). TW akcnowledges financial support from the Alfred Kordelin foundation. This project has received funding from the European Research Council (ERC) under the European Union's Horizon 2020 research and innovation programme (grant agreement no.~818665 ``UniSDyn''). The automated astronomy program at Tennessee State University has been supported by NASA, NSF, TSU and the State of Tennessee through the Centers of Excellence program. The authors thank L\'ucia Duarte for helpful discussions about the model--observation comparisons.
\end{acknowledgements}

\bibliographystyle{aa}
\bibliography{paper}

\begin{appendix}

\section{Additional data}

\begin{table*}
\centering
\caption{List of individual spectropolarimetric observations.}
\begin{tabular}{llccccrr@{$\,\pm\,$}l}
\hline
\hline
Instrument & Date & HJD & $\phi$\tablefootmark{a} & $(S/N)_I$\tablefootmark{b} & $(S/N)_V$\tablefootmark{c} & ${\rm FAP}$\tablefootmark{d} & \multicolumn{2}{c}{$\langle B_z \rangle$} \\
& (UTC) & (2\,450\,000+) & & & & & \multicolumn{2}{c}{[G]} \\
\hline
HARPSpol & 2010 Jan 5  & 5201.808 & 0.340 & 99.8  & 12775 & $<10^{-16}$ & $-75.4$&$6.6$ \\
         & 2010 Jan 6  & 5202.826 & 0.976 & 168.9 & 21982 & $<10^{-16}$ & $-11.9$&$3.9$ \\
         & 2010 Jan 7  & 5203.791 & 0.579 & 143.3 & 18337 & $<10^{-16}$ & $-38.4$&$4.6$ \\
         & 2010 Jan 8  & 5204.870 & 0.253 & 156.5 & 20156 & $<10^{-16}$ & $-55.1$&$4.2$ \\
         & 2010 Jan 9  & 5205.838 & 0.857 & 188.5 & 24264 & $<10^{-16}$ & $-53.6$&$3.5$ \\
         & 2010 Jan 10 & 5206.846 & 0.487 & 161.3 & 20723 & $<10^{-16}$ & $-16.8$&$4.1$ \\
         & 2010 Jan 13 & 5209.852 & 0.364 & 122.6 & 15522 & $<10^{-16}$ & $-60.6$&$5.4$ \\
         & 2010 Jan 14 & 5210.838 & 0.980 & 138.5 & 17543 & $<10^{-16}$ & $-20.6$&$4.8$ \\
         & 2010 Jan 15 & 5211.815 & 0.590 & 186.0 & 23814 & $<10^{-16}$ & $-24.7$&$3.5$ \\
         & 2010 Jan 16 & 5212.770 & 0.187 & 167.4 & 21421 & $<10^{-16}$ & $-5.3$&$4.0$ \\
\hline
HARPSpol & 2011 Feb 8  & 5600.662 & 0.447 & 136.9 & 16468 & $<10^{-16}$ & $-12.0$&$4.7$ \\
         & 2011 Feb 8  & 5600.789 & 0.526 & 153.2 & 18060 & $<10^{-16}$ & $-11.5$&$4.3$ \\
         & 2011 Feb 9  & 5601.632 & 0.053 & 126.9 & 15068 & $<10^{-16}$ & $21.0$&$5.1$ \\
         & 2011 Feb 9  & 5601.826 & 0.174 & 152.2 & 17974 & $<10^{-16}$ & $-6.1$&$4.3$ \\
         & 2011 Feb 10 & 5602.634 & 0.679 & 142.0 & 17160 & $<10^{-16}$ & $-22.4$&$4.5$ \\
         & 2011 Feb 10 & 5602.813 & 0.791 & 143.1 & 17043 & $<10^{-16}$ & $-4.2$&$4.5$ \\
         & 2011 Feb 12 & 5604.678 & 0.955 & 120.0 & 14376 & $<10^{-16}$ & $50.9$&$5.4$ \\
         & 2011 Feb 12 & 5604.843 & 0.058 & 106.8 & 12827 & $<10^{-16}$ & $21.2$&$6.0$ \\
         & 2011 Feb 13 & 5605.627 & 0.548 & 146.6 & 17658 & $<10^{-16}$ & $-9.0$&$4.4$ \\
         & 2011 Feb 13 & 5605.838 & 0.680 & 148.3 & 17670 & $<10^{-16}$ & $-31.3$&$4.4$ \\
         & 2011 Feb 14 & 5606.580 & 0.143 & 128.4 & 15375 & $<10^{-16}$ & $-6.9$&$5.0$ \\
         & 2011 Feb 14 & 5606.739 & 0.243 & 144.2 & 17101 & $<10^{-16}$ & $-26.7$&$4.5$ \\
         & 2011 Feb 15 & 5607.627 & 0.797 & 134.1 & 16178 & $<10^{-16}$ & $-7.3$&$4.8$ \\
         & 2011 Feb 15 & 5607.745 & 0.871 & 166.5 & 19834 & $<10^{-16}$ & $17.6$&$3.9$ \\
\hline
ESPaDOnS & 2016 Jan 14 & 7402.061 & 0.523 & 216.6 & 14996 & $3.1\times10^{-10}$ & $-25.6$&$7.6$ \\
         & 2016 Jan 15 & 7403.016 & 0.119 & 223.5 & 15286 & $1.7\times10^{-16}$ & $-42.3$&$7.4$ \\
         & 2016 Jan 16 & 7404.051 & 0.766 & 219.7 & 15271 & $4.9\times10^{-10}$ & $2.8$&$7.4$ \\
         & 2016 Jan 17 & 7405.048 & 0.388 & 233.7 & 16416 & $<10^{-16}$ & $-39.6$&$7.0$ \\
         & 2016 Jan 18 & 7406.044 & 0.010 & 230.3 & 15829 & $<10^{-16}$ & $-13.6$&$7.0$ \\
         & 2016 Jan 19 & 7407.076 & 0.655 & 231.8 & 15851 & $7.7\times10^{-11}$ & $-19.1$&$7.1$ \\
         & 2016 Jan 21 & 7409.038 & 0.880 & 177.9 & 12077 & $<10^{-16}$ & $32.9$&$9.2$ \\
         & 2016 Jan 22 & 7410.042 & 0.507 & 196.2 & 13353 & $5.1\times10^{-13}$ & $-44.3$&$8.6$ \\
         & 2016 Jan 23 & 7411.006 & 0.109 & 245.1 & 16931 & $<10^{-16}$ & $-41.8$&$6.6$ \\
\hline
HARPSpol & 2017 Dec 13 & 8100.823 & 0.939 & 198.9 & 15661 & $<10^{-16}$ & $33.2$&$3.2$ \\
         & 2017 Dec 14 & 8101.835 & 0.571 & 203.9 & 16221 & $<10^{-16}$ & $-32.9$&$3.1$ \\
         & 2017 Dec 15 & 8102.833 & 0.194 & 218.0 & 17400 & $<10^{-16}$ & $-33.0$&$2.9$ \\
         & 2017 Dec 16 & 8103.833 & 0.819 & 202.0 & 15892 & $<10^{-16}$ & $56.0$&$3.1$ \\
         & 2017 Dec 17 & 8104.701 & 0.361 & 184.1 & 14539 & $<10^{-16}$ & $1.0$&$3.5$ \\
         & 2017 Dec 17 & 8104.822 & 0.437 & 181.9 & 14485 & $<10^{-16}$ & $5.0$&$3.5$ \\
         & 2017 Dec 17 & 8104.852 & 0.455 & 141.9 & 11214 & $<10^{-16}$ & $4.4$&$4.5$ \\
         & 2017 Dec 19 & 8106.827 & 0.689 & 95.7  & 7532  & $<10^{-16}$ & $-0.1$&$6.7$ \\
         & 2017 Dec 19 & 8106.857 & 0.708 & 85.5  & 6886  & $<10^{-16}$ & $25.6$&$7.2$ \\
\hline
\end{tabular}
\tablefoot{
\tablefoottext{a}{Phases $\phi$ calculated using the ephemeris of Eq.~\ref{eq-epoch};}
\tablefoottext{b}{$S/N$ of the Stokes $I$ spectrum;}
\tablefoottext{c}{$S/N$ of the Stokes $V$ LSD profile;}
\tablefoottext{d}{${\rm FAP}$ of the magnetic field detection in the Stokes $V$ LSD profile}
}
\label{tab-obs}
\end{table*}

\begin{table*}
\centering
\caption{List of chromospheric activity indices for individual observations.}
\begin{tabular}{llccccc}
\hline
\hline
Instrument & Date & HJD & $\phi$\tablefootmark{a} & $S_{\rm MWO}$ & $\log R'_{\rm HK}$ \\
& (UTC) & (2\,450\,000+) & & \\
\hline
HARPSpol & 2010 Jan 5  & 5201.808 & 0.340 & 2.262 & -3.836 \\
         & 2010 Jan 6  & 5202.826 & 0.976 & 2.014 & -3.887 \\
         & 2010 Jan 7  & 5203.791 & 0.579 & 1.920 & -3.908 \\
         & 2010 Jan 8  & 5204.870 & 0.253 & 1.970 & -3.897 \\
         & 2010 Jan 9  & 5205.838 & 0.857 & 2.104 & -3.868 \\
         & 2010 Jan 10 & 5206.846 & 0.487 & 2.051 & -3.879 \\
         & 2010 Jan 13 & 5209.852 & 0.364 & 1.930 & -3.906 \\
         & 2010 Jan 14 & 5210.838 & 0.980 & 1.911 & -3.911 \\
         & 2010 Jan 15 & 5211.815 & 0.590 & 2.074 & -3.874 \\
         & 2010 Jan 16 & 5212.770 & 0.187 & 2.016 & -3.887 \\
\hline
HARPSpol & 2011 Feb 8  & 5600.662 & 0.447 & 2.093 & -3.870 \\
         & 2011 Feb 8  & 5600.789 & 0.526 & 1.997 & -3.891 \\
         & 2011 Feb 9  & 5601.632 & 0.053 & 2.037 & -3.882 \\
         & 2011 Feb 9  & 5601.826 & 0.174 & 1.958 & -3.900 \\
         & 2011 Feb 10 & 5602.634 & 0.679 & 2.108 & -3.867 \\
         & 2011 Feb 10 & 5602.813 & 0.791 & 2.267 & -3.835 \\
         & 2011 Feb 12 & 5604.678 & 0.955 & 2.115 & -3.866 \\
         & 2011 Feb 12 & 5604.843 & 0.058 & 1.921 & -3.908 \\
         & 2011 Feb 13 & 5605.627 & 0.548 & 1.910 & -3.911 \\
         & 2011 Feb 13 & 5605.838 & 0.680 & 2.123 & -3.864 \\
         & 2011 Feb 14 & 5606.580 & 0.143 & 2.185 & -3.851 \\
         & 2011 Feb 14 & 5606.739 & 0.243 & 2.079 & -3.873 \\
         & 2011 Feb 15 & 5607.627 & 0.797 & 2.061 & -3.877 \\
         & 2011 Feb 15 & 5607.745 & 0.871 & 2.004 & -3.889 \\
\hline
ESPaDOnS & 2016 Jan 14 & 7402.061 & 0.523 & 2.120 & -3.865 \\
         & 2016 Jan 15 & 7403.016 & 0.119 & 2.232 & -3.842 \\
         & 2016 Jan 16 & 7404.051 & 0.766 & 2.335 & -3.822 \\
         & 2016 Jan 17 & 7405.048 & 0.388 & 1.945 & -3.903 \\
         & 2016 Jan 18 & 7406.044 & 0.010 & 2.298 & -3.829 \\
         & 2016 Jan 19 & 7407.076 & 0.655 & 2.131 & -3.862 \\
         & 2016 Jan 21 & 7409.038 & 0.880 & 2.182 & -3.852 \\
         & 2016 Jan 22 & 7410.042 & 0.507 & 1.797 & -3.938 \\
         & 2016 Jan 23 & 7411.006 & 0.109 & 2.070 & -3.875 \\
\hline
HARPSpol & 2017 Dec 13 & 8100.823 & 0.939 & 2.080 & -3.873 \\
         & 2017 Dec 14 & 8101.835 & 0.571 & 2.106 & -3.867 \\
         & 2017 Dec 15 & 8102.833 & 0.194 & 2.025 & -3.885 \\
         & 2017 Dec 16 & 8103.833 & 0.819 & 2.248 & -3.838 \\
         & 2017 Dec 17 & 8104.701 & 0.361 & 2.133 & -3.862 \\
         & 2017 Dec 17 & 8104.822 & 0.437 & 2.197 & -3.849 \\
         & 2017 Dec 17 & 8104.852 & 0.455 & 2.117 & -3.865 \\
         & 2017 Dec 19 & 8106.827 & 0.689 & 2.105 & -3.868 \\
         & 2017 Dec 19 & 8106.857 & 0.708 & 2.208 & -3.846 \\
\hline
\end{tabular}
\tablefoot{
\tablefoottext{a}{Phases $\phi$ calculated using the ephemeris of Eq.~\ref{eq-epoch};}
}
\label{tab-chrom}
\end{table*}

\begin{figure*}[h]
\sidecaption
\begin{tabular}{cc}
\includegraphics[bb=40 65 355 775,width=5.5cm,clip]{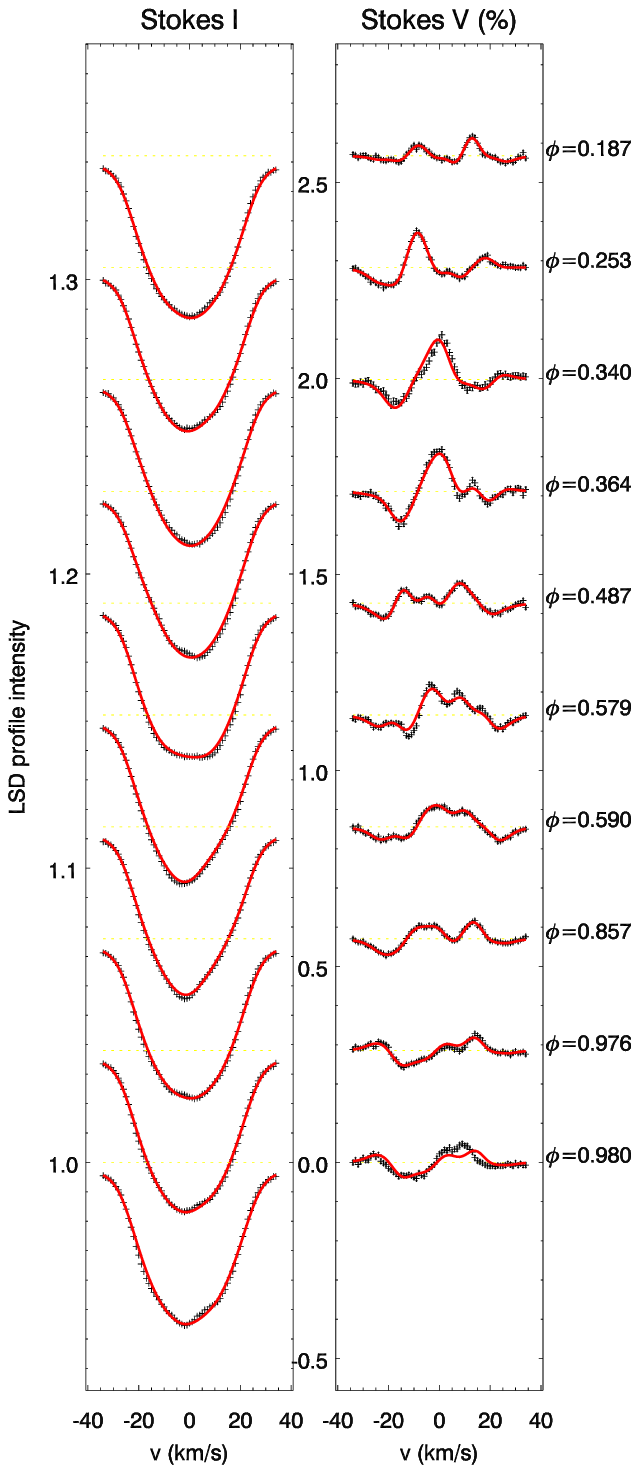} &
\includegraphics[bb=40 65 355 775,width=5.5cm,clip]{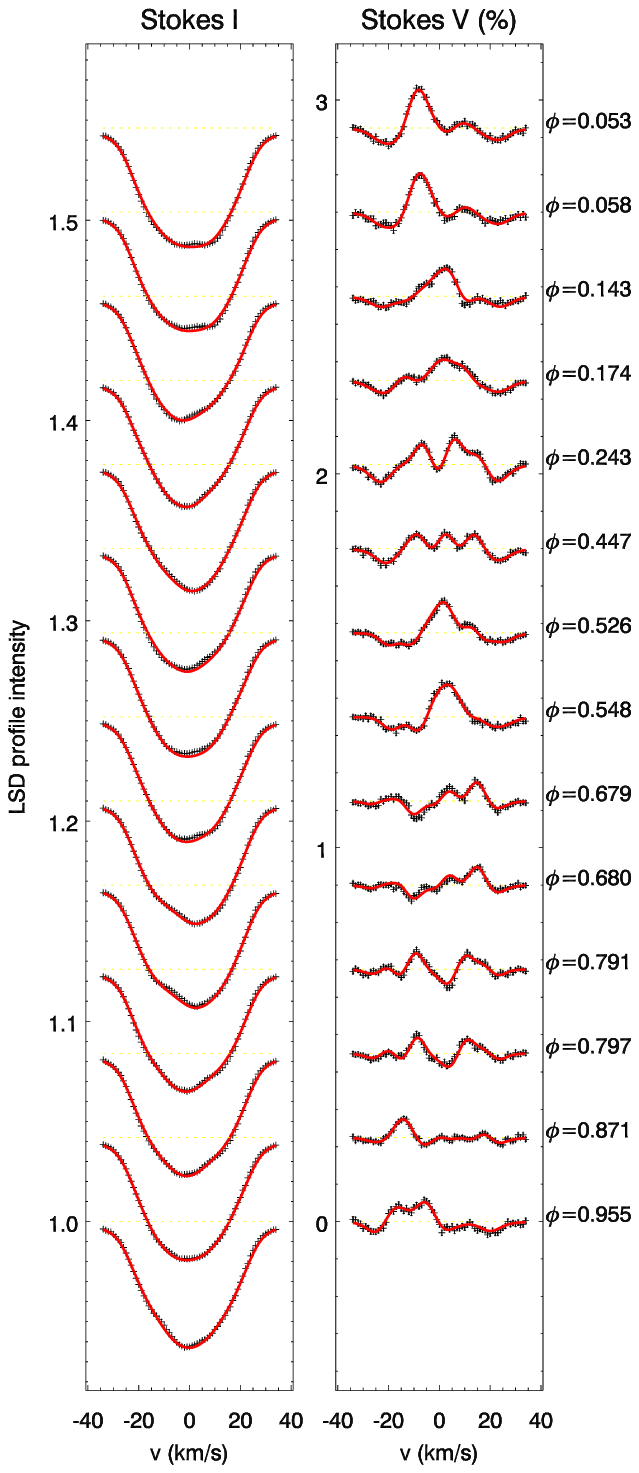} \\
\includegraphics[bb=40 65 355 775,width=5.5cm,clip]{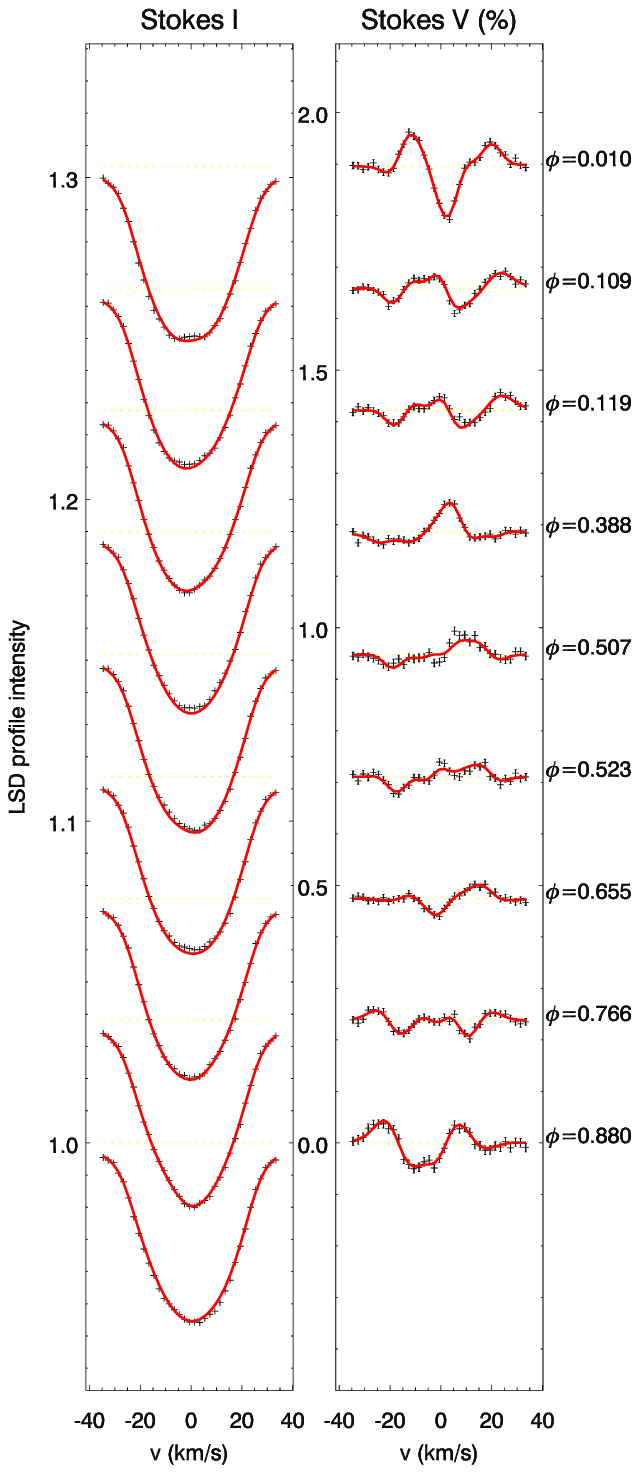} &
\includegraphics[bb=40 65 355 775,width=5.5cm,clip]{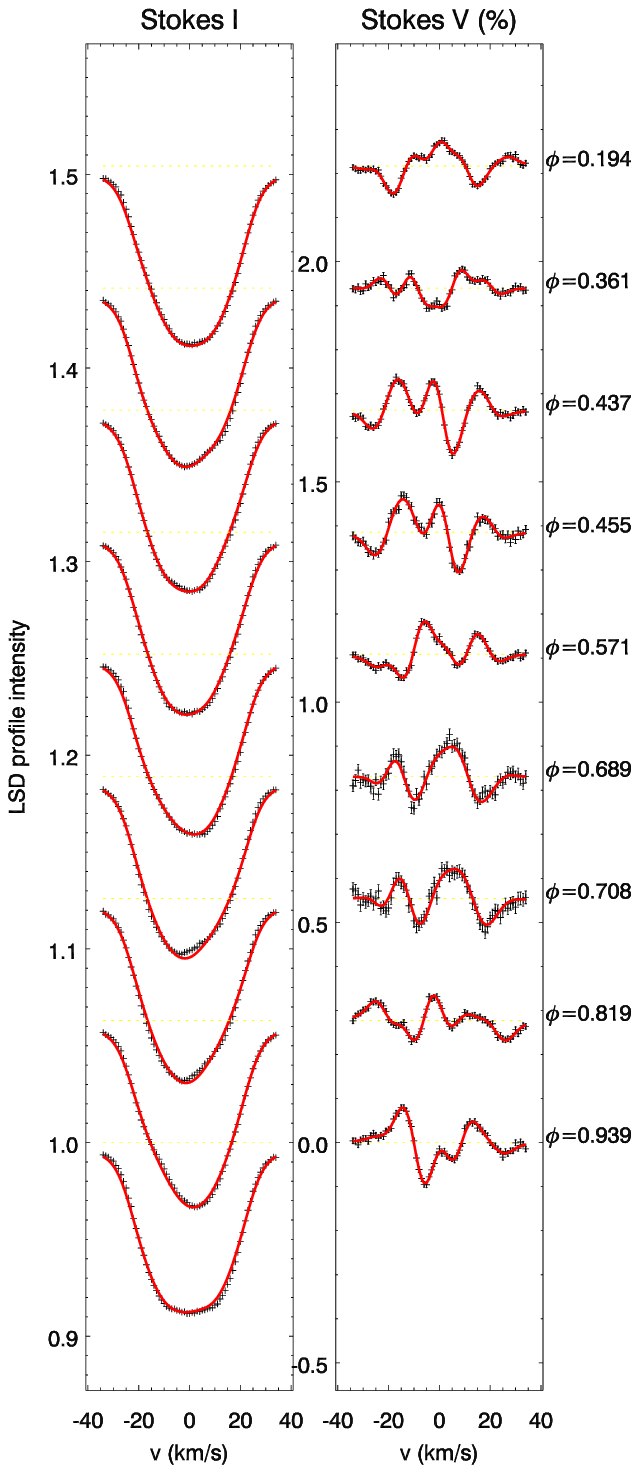}
\end{tabular}
\caption{Stokes $I$ and $V$ LSD profiles (black) and the line profile fits (red) for the epochs January 2010 (\textit{top left}), February 2011 (\textit{top right}), January 2016 (\textit{bottom left}), and December 2017 (\textit{bottom right}). The observed profiles are shifted vertically for clarity and their rotational phases, $\phi$, are given next to each profile.}
\label{fig-prof}
\end{figure*}

\end{appendix}

\end{document}